\documentclass[aps,prb,superscriptaddress,twocolumn,tightenlines,showpacs,floatfix,amsmath,amssymb,pdftex]{revtex4-2}
\usepackage{graphicx}
\usepackage{xcolor}
\usepackage{physics}
\usepackage[colorlinks=true,urlcolor=blue,citecolor=blue,allcolors=blue]{hyperref}
\usepackage{amsfonts}
\usepackage{amssymb}
\usepackage{bbm}
\DeclareMathOperator*{\Motimes}{\text{\raisebox{0.25ex}{\scalebox{0.8}{$\bigotimes$}}}}

\renewcommand{\vec}{\mathbf}

\definecolor{purple}{rgb}{0.7, 0., 0.8}

\begin{document}

\title{Thermalization by a synthetic horizon}

\author{Lotte Mertens}\email{l.mertens@uva.nl}
\affiliation{Institute for Theoretical Physics Amsterdam,
University of Amsterdam, Science Park 904, 1098 XH Amsterdam, The Netherlands}
\affiliation{Institute for Theoretical Solid State Physics, IFW Dresden, Helmholtzstr. 20, 01069 Dresden, Germany}
\author{Ali G. Moghaddam}\email{agorbanz@iasbs.ac.ir}
\affiliation{Department of Physics, Institute for Advanced Studies in Basic Sciences (IASBS), Zanjan 45137-66731, Iran}
\affiliation{Computational Physics Laboratory, Physics Unit, Faculty of Engineering and
Natural Sciences, Tampere University, FI-33014 Tampere, Finland}
\affiliation{Institute for Theoretical Solid State Physics, IFW Dresden, Helmholtzstr. 20, 01069 Dresden, Germany}
\author{Dmitry Chernyavsky}
\affiliation{Institute for Theoretical Solid State Physics, IFW Dresden, Helmholtzstr. 20, 01069 Dresden, Germany}
\author{Corentin Morice}
\affiliation{Institute for Theoretical Physics Amsterdam,
University of Amsterdam, Science Park 904, 1098 XH Amsterdam, The Netherlands}
\affiliation{Laboratoire de Physique des Solides, CNRS UMR 8502, Université Paris-Saclay, F-91405
Orsay Cedex, France}
\author{Jeroen van den Brink}
\affiliation{Institute for Theoretical Solid State Physics, IFW Dresden, Helmholtzstr. 20, 01069 Dresden, Germany}
\affiliation{Institute for Theoretical Physics and Würzburg-Dresden Cluster of Excellence ct.qmat, Technische Universität Dresden, 01069 Dresden, Germany}
\author{Jasper van Wezel}
\affiliation{Institute for Theoretical Physics Amsterdam,
University of Amsterdam, Science Park 904, 1098 XH Amsterdam, The Netherlands}

\date{\today}

\begin{abstract}
Synthetic horizons in models for quantum matter provide an alternative route to explore fundamental questions of modern gravitational theory.
Here, we apply these concepts to the problem of emergence of thermal quantum states in the presence of a horizon, by studying ground-state thermalization due to instantaneous horizon creation in a gravitational setting and its condensed matter analogue. 
By a sudden quench to position-dependent hopping amplitudes in a one-dimensional lattice model, we establish the emergence of a thermal state accompanying the formation of a synthetic horizon. The resulting temperature for long chains is shown to be identical to the corresponding Unruh temperature,
provided that the post-quench Hamiltonian matches the entanglement Hamiltonian of the pre-quench system.
Based on detailed analysis of the outgoing radiation we formulate the conditions required for the synthetic horizon to behave as a purely thermal source, paving a way to explore this interplay of quantum-mechanical and gravitational aspects experimentally.
\end{abstract}

\maketitle

\section{Introduction} Ever since Unruh introduced the sonic black hole \cite{unruh1981}, the   promise of controllable experimental access to the physics of general relativity has driven searches for condensed matter analogues of gravitational systems, e.g., in the context of optical and magnonic systems \cite{philbin2008,duine2017}, classical electronic circuits \cite{kollar2019hyperbolic,Boettcher2020,Boettcher2022}, superfluids \cite{hu2019Unruh,Nissinen2020,Barcelo2001,barcelo2011analogue,Carusotto2008,nguyen2015}, and fermionic systems \cite{Calabrese2017,Minar2015,Jafari2019, staalhammar2021pt}.
Here, we investigate the thermalization of an electronic quantum system with a synthetic horizon~\cite{kedem2020black,morice2021prr,Volovik:2016kid,Sabsovich2022,DeBeule2021}, which in principle admits straightforward experimental implementation~\cite{eigler_schweizer_1990, hirjibehedin_2006, drost_ojanen_harju_liljeroth_2017}. 
 
Thermalization in gravity is well-known to arise already in flat spacetime, where static observers can detect the pure vacuum state, while accelerated observers in the same spacetime will have a horizon making part of the ground state unobservable to them~\cite{Unruh1976}. The vacuum then appears mixed to an accelerated observer, and in fact turns out to look thermal~\cite{takagi_1986}. The same happens to a static observer near a black hole horizon~\cite{Srednicki1993,Sorkin1986,carroll_2020}. This raises the question how  a quantum ground state thermalizes as a horizon emerges during black hole formation and spawns the associated problem of information loss paradox~\cite{Hawking1976,Page2005,susskind2006paradox}.

\begin{figure}
    \centering
    \includegraphics[width=\linewidth]{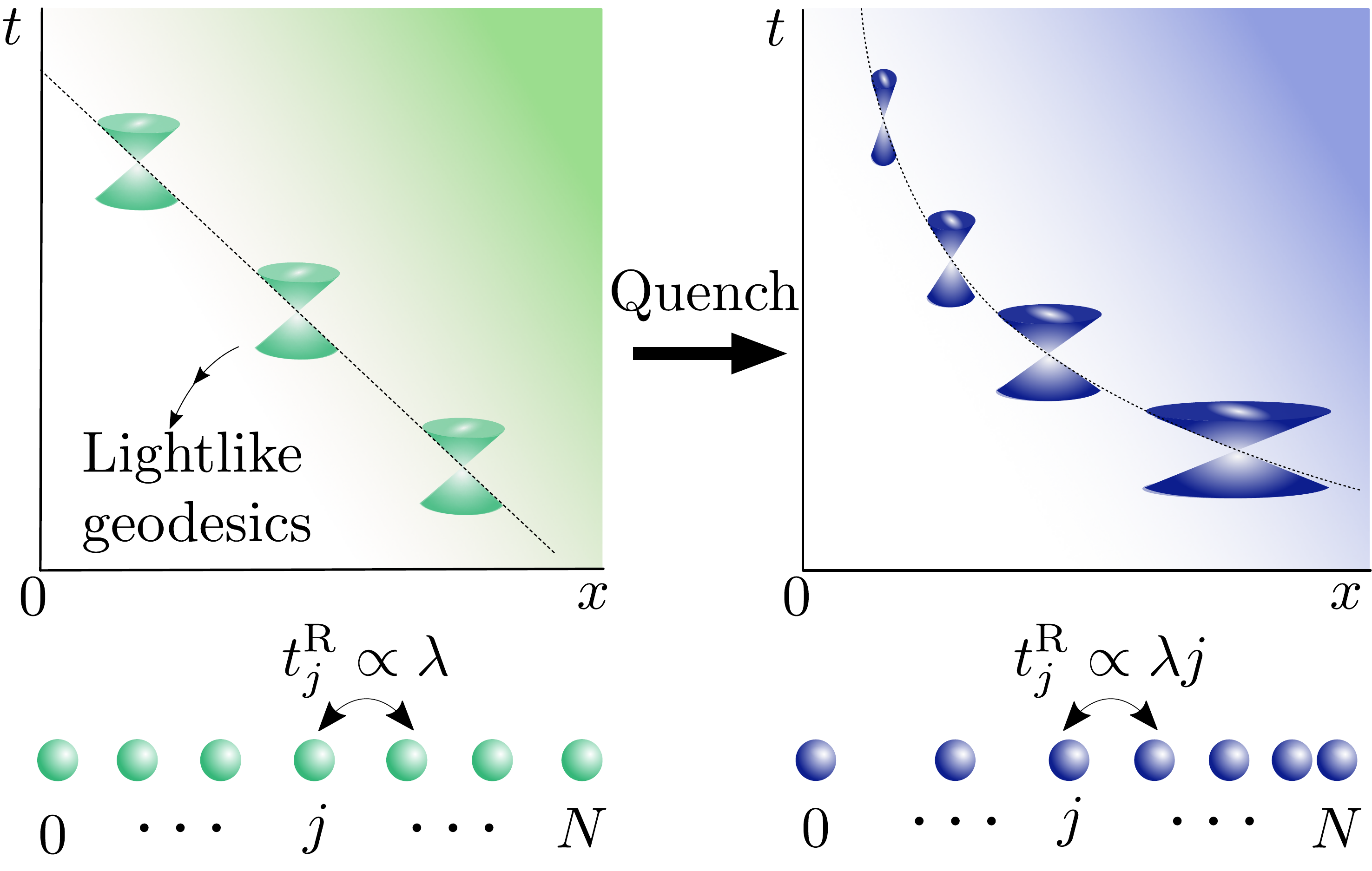}
    \caption{Schematic overview of the quench set-up. In the top part of the figure a lightlike geodesic is drawn in a spacetime diagram for flat spacetime (left) and curved spacetime with a horizon at $x=0$ (right). In the lower part the corresponding tight binding models are shown with constant hopping (left) and position dependent hopping (right).}
    \label{fig:quench}
\end{figure}

As the process of black hole formation is in general complex and highly sensitive to the gravitational collapse dynamics, we here employ a minimal theoretical model that retains only its key feature: a non-singular initial configuration collapsing into a singular end state with a horizon. We focus on a (1+1)D setting and consider a horizon forming at position $x=0$. We presume this formation to happen as an {\it instantaneous} quench at $t=0$, effectively neglecting all detail of the collapse dynamics as illustrated in Fig. \ref{fig:quench}. The effect of the quench is characterized by calculating the number of particles measured by a static observer before and after formation of the horizon.
We then also introduce a condensed matter analogue in the form of a one-dimensional tight-binding model as visible in the lower part of Fig. \ref{fig:quench}. A synthetic horizon in this analogue is created there by a quench of the homogeneous system into one with particular position-dependent hopping parameters~\cite{morice2021prr}.
The low-energy properties of the lattice model after the quench are given by the Dirac equation on a (1+1)D black hole background as it arises in Jackiw-Teitelboim (JT) gravity.
We show that the quench results in an emergent temperature in the lattice model equal to the Unruh temperature in the gravitational system, provided that: ($i$) the chain consists of a large but finite number of sites, and ($ii$) the post-quench Hamiltonian corresponds to the entanglement Hamiltonian of the half-system bipartition of the system before the quench. Under these conditions we find that the horizon behaves as a purely thermal source. 

\section{Quench in General Relativity}
Let us start from flat space ($ds^2 = dT^2 - dX^2$) with a many-particle ground state or vacuum field configuration $\ket{0_{\text{M}}}$ in which all negative energy modes are filled. Switching from an inertial to an accelerated observer is described by the coordinate transformation
\begin{align*}
    T = x \sinh(\Gamma t), \  
    X = x \cosh(\Gamma t).
\end{align*}
The spacetime in terms of proper coordinates $x,t$ can be described by the Rindler metric\cite{carroll_2020, misner_thorne_wheeler_2002} $ds^2 = \Gamma^2 x^2 dt^2 - dx^2$. By the equivalence principle, any casual horizon can be approximated by the Rindler metric in a small region of space-time, such as the spacetime structure close to a Schwarzschild black hole horizon. 

We anticipate that in the lattice analogy it will not be possible to switch between different observers due to the physical reference frame provided by the atomic background. In the current relativistic setting we therefore also insist on having a single observer, and we interpret $X,T$ and $x,t$ to be proper coordinates for the \emph{same} observer before and after a quench that changes spacetime. The coordinates coincide at the event $x=X=X_0, t=T=0$ and observers at all positions agree about the time $t=T=0$, which we take to be the time at which the quench is enacted. At that moment, the observer instantaneously goes from being inertial to having the acceleration $\alpha = \frac{1}{\Gamma x_0}$, which we interpret as being due to the instantaneous appearance of a gravitational mass in spacetime.

Immediately after the quench, a static detector (unchanging $x$-coordinate and thus not free-falling) will measure particles defined with respect to the Rindler metric. We denote their creation operators as $\hat{b}_{E_\text{R}}^{(1)\dagger}$, with $\epsilon$ the energy of the state and the $(1)$ indicating that the state is created to the right of the horizon in Rindler spacetime. Using the standard approach involving local transformations and the overlap matrix between states defined in Rindler and Minkowski (flat) spacetimes, the vacuum expectation value for the number of particles observed by a static detector at positive times becomes 
\begin{align}
        \bra{0_{\text{M}}}\hat{b}^{(1)\dagger}_{E_\text{R}} \hat{b}^{(1)}_{{E_\text{R}}'}\ket{0_{\text{M}}}  =  \frac{1}{1+ e^{2 \pi {E_\text{R}} /\Gamma}} \delta({E_\text{R}}-{E_\text{R}}'). \label{eq:FDGR1}
\end{align}
Here, $\ket{0_{\text{M}}}$ is the Minkowski vacuum. We recognize this as the Fermi-Dirac distribution for particles in thermal equilibrium at temperature $T = \Gamma/(2\pi)$. The detailed derivation of the geodesics and wavefunctions on the Minkowski and Rindler metrics and the calculation of the thermal spectrum in general relativity are presented in Appendix \ref{sec:grav}.

The instantaneous creation at $t=0$ of a horizon at $x=0$ by quenching from flat spacetime to the Rindler metric caused the definition of particles to change. Concomitantly, this causes static observers to detect a thermal distribution of particles rather than the zero-temperature vacuum. The resulting temperature is a quantity akin to Hawking and Unruh radiation, in the sense that it emerges due to the presence of the horizon. We now proceed to establish a condensed matter analogue for this effect and the corresponding emerging temperature at a synthetic horizon.

\section{Synthetic horizon}
It has previously been established that tight-binding models with position-dependent hopping can exhibit synthetic horizons~\cite{morice2021prr}. Briefly summarizing the central results, we note that a general non-interacting model with nearest-neighbor hopping can be written as

\begin{align}
\label{eq:H}
    \hat{H} = \sum_{j=1}^{N/2-1} t_j^{\text{L}} \hat{c}_j^\dagger \hat{c}_{j+1}^{\phantom \dagger} + t_j^{\text{R}} \hat{c}_{j+N/2}^\dagger \hat{c}_{j + 1 + N/2}^{\phantom \dagger} + \text{H.c.}, 
\end{align}
where $j$ labels lattice sites. We define position-dependent hopping amplitudes to the right of $j=N/2$ as
\begin{align*}
    t_j^{\text{R}} = \lambda \Big(\frac{ j }{N/2-1}\Big)^{\gamma}.
\end{align*}
For constant hopping, $\gamma=0$, and at half filling the group velocity of a wave packet, $v = \lambda$, is constant throughout the lattice and defines wave packet trajectories. These trajectories in the analogue model coincide with the geodesics of light in flat spacetime. For $\gamma \neq 0$, the hopping amplitude depends on position and naively gives rise to a locally varying group velocity. At the middle of the chain the hopping amplitude vanishes entirely, and the local group velocity likewise goes to zero. Depending on $\gamma$ this causes the formation of a synthetic horizon with wave packets approaching the center of the chain slowing down and never reaching it exactly. As it has been shown in Ref. \cite{morice2021prr}, considering a power-law profile for the hopping amplitude as $t~j^\gamma$
only for $\gamma \geq 1$, an analog horizon exists.

This behavior is manifested in the low-energy effective description of the system in terms of a Dirac equation by its position dependent Dirac cones~\cite{morice2021prr}. Focusing on the special case of $\gamma=1$ from here on, the wave packet trajectories at low energy are given by~\cite{morice2021prr}
\begin{align*}
    x(\tau) = x_0 e^{\pm 2 \lambda  \tau /(N/2-1)}.
\end{align*}
Here, $x$ is the continuum position that becomes $x=j-N/2$ at lattice sites. Only at $\tau \pm \infty $ does the wave packet reach the middle of the lattice at $x=0$. The trajectory in the lattice model moreover coincides with the geodesics of light in the Rindler metric with $\Gamma = 2\lambda/(N/2-1)$. Only for the special $\gamma =1$ case these world lines of wave packets on the lattice are mapped onto those of massless particles in a gravitational black hole metric. If $N/2$ is odd one can obtain the exact eigenfunction of the Hamiltonian for the zero mode with energy $E_{\text{R}}=0$. Modes with non-zero energy are well approximated at low energies and large $N$ by~\cite{ali}
\begin{align}
    \phi_n = \frac{1}{\sqrt{2\pi}}i^{n} n^{-1/2+i\frac{E_{\text{R}}}{2\lambda}(N/2-1)} \label{eq:continuous}.
\end{align}
These modes match particle wave functions defined in the background of the continuous gravitational system for $\Gamma = 2\lambda/(N/2-1)$, up to the factor $i^n$. In appendix \ref{sec:AppComp} we present more details about the comparison between both the wavefunctions, geodesics and energy spectra for the tight binding models to their general relativity analogues as well as the overlap function to the thermal spectrum. 

\section{Quench and emergent temperature of lattice system}
\begin{figure}[tb]
    \centering
    \includegraphics[width=\linewidth]{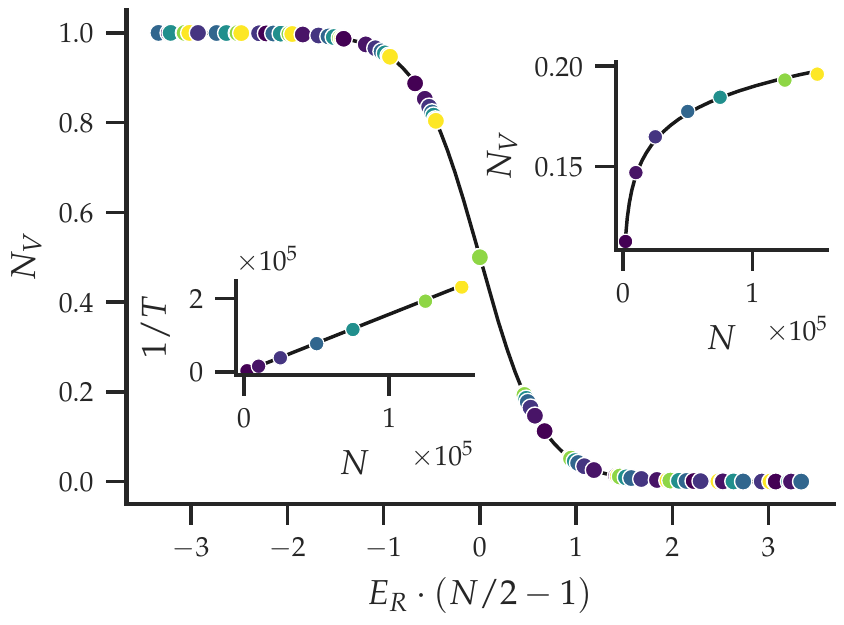}
    \caption{Diagonal elements of the overlap matrix $N_V$, calculated numerically for system sizes ranging between $N=2006$ (dark blue), through $N=20006$, $25006$, $50006$, $75006$, $100006$, and $125006$, to $N=150006$ (yellow). To enable comparison between different system sizes, $N_V$ is shown as a function of energy times the number of sites , $E_\text{R}(N/2 -1)$ such that it gives a constant temperature. The thermal Fermi-Dirac distribution of Eq.~\eqref{eq:LatticeNv} (black line) can be clearly distinguished. The right inset shows the scaling of $N_V$  with system size $N$ for the first energy level above zero. The black line is the best fit $N_V = 0.019 \log(N) - 0.032$.  While the left inset indicates the linear scaling of $\beta=1/T$ with system size $N$, the black line is the continuous expectation $1/T = \frac{ \lambda(N/2-1)}{\pi}$. }
    \label{fig:ThermalPoints}
\end{figure}

For the gravitational quench we established the presence of a thermal spectrum given by equation~\eqref{eq:FDGR1}. The analogous quantity in the lattice model of equation~\eqref{eq:H} is $ N_V \equiv \bra{\text{GS}_{\text{M}}} \hat{b}_{E_{\text{R}}}^\dagger \hat{b}_{{\bar{E}_{\text{R}}}}^{\phantom \dagger} \ket{\text{GS}_{\text{M}}}$, where $\ket{\text{GS}_{\text{M}}}$ is the ground state with constant hopping ($\gamma=0$) and for half-filling with $N/2$ occupied states~\footnote{We focused on half-filling but this is not crucial for the argument, we could adjust the method for arbitrary partial-filling state, too }.
$\hat{b}_{E_{\text{R}}}^\dagger$ creates an eigenstate of energy $E_{\text{R}}$ in the right side of the system with linearly increasing hopping amplitude ($\gamma=1$). This expression can be written in a more intuitive fashion using the overlap functions on the lattice
\begin{align}
    V_{E_{\text{M}},{E_{\text{R}}}} &= \braket{E_{\text{M}}}{E_{\text{R}}} = \sum_j \bra{E_{\text{M}}}\ket{j}\bra{j}\ket{E_{\text{R}}} \notag \\
    \Rightarrow ~~~ N_V &=  \sum_{E_{\text{M}}<0} V_{E_{\text{M}},{E_{\text{R}}}}V_{E_{\text{M}}, {\bar{E}_{\text{R}}}}
    \label{eq:Vdef}
\end{align}
The overlap $N_V$ is numerically verified in Fig.~\ref{fig:ThermalPoints} to obtain the thermal shape
\begin{align}
\label{eq:LatticeNv}
    N_V({E_{\text{R}}}, {\bar{E}_{\text{R}}}) = \delta_{{E_{\text{R}}}, {\bar{E}_{\text{R}}}} \frac{1}{1 + e^{\pi / \lambda {E_{\text{R}}}(N/2-1)}}.
\end{align}

As all off-diagonal elements of the overlap matrix $N_V$ are found to vanish, the main plot in Fig.~\ref{fig:ThermalPoints} displays the diagonal elements with $\bar{E}_\text{R}=E_\text{R}$. Plotting $N_V$ as function of energy times the number of sites $(N/2-1)$, such that curves for different system sizes coincide and the spectrum can be seen to become continuous in the thermodynamic limit $N \to \infty$. The curve exactly follows a Fermi-Dirac distribution with temperature scaling linearly with system size and reproducing the continuum expectation $T^{-1} = (\pi/\lambda)(N/2-1) \equiv 2\pi/\Gamma$. This temperature dependence can be seen in the left inset, where the black line is the relation $T^{-1} = (\pi/\lambda)(N/2-1) \equiv 2\pi/\Gamma$. In the thermodynamic limit ($N \to \infty$) the temperature goes to zero and the Fermi-Dirac function becomes a step function in which all states with negative energies are occupied, and the states with positive energies are empty. This can be understood because we take the limit of $N$ to infinity while keeping the maximum hopping amplitude (at the final bond) constant. In this choice for the definition of the continuum limit, the difference between hopping amplitudes on neighboring sites goes to zero, which implies that acceleration, and therefore temperature in the lattice model, also go to zero. 

Equivalently, the temperature can be said to vanish in the thermodynamic limit in the lattice model because it obtains a locally constant hopping everywhere, while in the corresponding gravitational analogue no more energy is gained by local moves through the gravitational potential. The right inset of Fig.~\ref{fig:ThermalPoints} shows the scaling of $N_V$ with system size $N$ for the first energy level above zero in the lattice model, the black line is the best fit $N_V=0.019\log(N)-0.032$ which shows that in the thermodynamic limit all points on the Fermi-Dirac distribution are covered. The resulting distribution originating from the quench is the instantaneous distribution directly following the quench and thus not predict any late time behavior.
 
\section{Origin of the thermal distribution} 
In the gravitational model a Bogoliubov transformation is well-known to lead to the emergence of a thermal distribution. Notice however that because the quench of the lattice system from $\gamma=0$ to $\gamma=1$ does not affect the number of electrons in the chain, going between states $\ket{E_{\text{M}}}$ and $\ket{E_{\text{R}}}$, therefore it cannot be a Bogoliubov transformation. We show here that the overlap function $N_V$ in the lattice model nevertheless reduces to a thermal distribution because the linear hopping Hamiltonian is of a special form and because the system is effectively cut into separate subsystems by the quench~\cite{Peschel2009}.

For a general free fermionic system  divided into disjoint parts, left (L) and right (R), an entanglement Hamiltonian can be defined as $h_{\text{R}} = \log(\rho_{\text{R}})$ with $\rho_{\text{R}}$ the reduced density matrix for the right subsystem. It has been shown~\cite{Peschel2009} that the operators $\hat{b}^\dagger_k$ diagonalizing $h_{\text{R}}$ always have correlations in the ground state of the full system of the form
\begin{align}
\label{eq:correlations_a1}
    \langle b_k^\dagger b_l^{\phantom \dagger} \rangle & = \frac{\delta_{k, l}} {e^{\epsilon_k} +1 },
\end{align}
where $\epsilon_k$ are the eigenvalues of the entanglement Hamiltonian (see Appendix \ref{sec:AppEntan} for the reproduction of the link between eigenstates of an entanglement Hamiltonian and the thermal spectrum).

The reduced density matrix for a subsystem consisting half of an open chain of size $N$ with constant hopping can be well approximated by~\cite{dalmonte2018quantum,peschel_2004}
\begin{align*}
    & \rho_{R} \approx \frac{1}{\cal A}\exp \left( -\pi \sum_{j=1}^{N/2-1} j \: c_j^\dag c_{j+1}^{\phantom \dagger}\right),
\end{align*}
with normalization constant ${\cal A}$. 
To leading order in $1/N$, this equals to a thermal system with a constant temperature $T=(\lambda/\pi)/(N/2 -1 )$ and a Hamiltonian equal to the linear hopping model of equation~\eqref{eq:H}. Therefore, we can then use the result of equation~\eqref{eq:correlations_a1} to obtain the Fermi-Dirac distribution in the lattice model with this temperature. This explains the thermal character of the correlations in the lattice model: the quenched system is approximately equal to the entanglement Hamiltonian of the original system. It is the entanglement between the two sides of the system in the ground state of the constant hopping model that causes the thermal distribution of quenched particles, rather than a Bogoliubov transformation as in the gravitational model. In other words, in the gravitational model the transformed nature of the elementary particles in different background curvatures gives rise to a thermal distribution, while in the lattice model the emergence of temperature is rooted in the transformed eigenstates of fundamental particles under different Hamiltonians. In both cases, the quench causes entanglement across a horizon, either as a thermofield double configuration of gravitational modes \cite{israel1976thermo,takahashi1996thermo,Chapman2019}, or in the many-body ground state of electrons on a lattice.

\begin{figure}[tb]
    \centering
    \includegraphics[width=\linewidth]{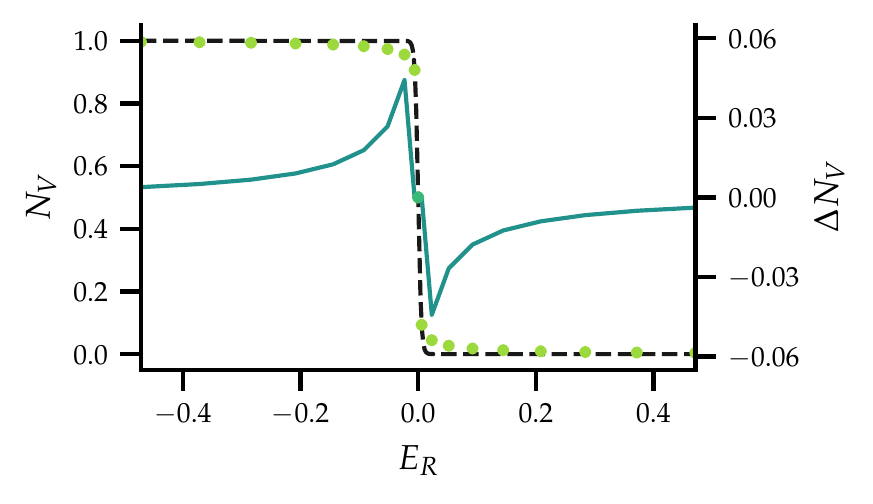}
    \caption{Diagonal elements of the overlap matrix $N_V$, calculated numerically for $N=1006$ after quenching to a model with $\gamma=2$. The black dashed line is the best fit to the numerical data using a Fermi-Dirac distribution, with $T=0.0025$. On the right axis the blue line denotes the difference between the best fit and the exact values, which diverges close to zero energy for large $N$.}
    \label{fig:FDgamma2}
\end{figure}

When quenching a lattice model with values of $\gamma$ different from one, such as the quadratic hopping model that still has a horizon~\cite{morice2021prr}, the quenched Hamiltonian is no longer proportional to an entanglement Hamiltonian of the original model,  and the correlations cease to be thermal. We confirmed this numerically, as shown in Fig. \ref{fig:FDgamma2}. Contrary to earlier statements~\cite{Rodriguez2017}, we thus find that the loss of information when quenching to a system with a horizon is not a sufficient condition for obtaining a thermal distribution. Starting from a many-body ground state on a lattice, a quench will yield precisely a thermal distribution of excited states {only if the new Hamiltonian is proportional to the entanglement Hamiltonian of the original system.}

\section{Summary and discussion}
We have presented a lattice analogue for the thermalization of the vacuum during the creation of the horizon. Despite having only a perfect overlap at low energies for the lattice and gravitational wave functions, as well as the gravitational geodesics and semi-classical wave packet trajectories on the lattice, an exact Fermi-Dirac thermal distribution emerges in both the lattice and gravitational models for all energies. That the thermal spectrum emerges independently of the wave functions agreeing, points to a common physical origin for the emergent temperature that is independent of the precise realization in terms of energy eigenstates, indicating the origin for temperature is the presence of horizons in both models.

Keeping the maximum hopping amplitude constant when varying the number of sites in the lattice model, the distance between points in the thermal overlap function $N_V$ was found to decrease logarithmically with system size. The thermal spectrum thus becomes continuous in the thermodynamic limit. At the same time, however, the obtained temperature goes down as $T \propto 1/N $, giving zero temperature in the thermodynamic limit \footnote{A similar property has been seen for an effective temperature defined from reduced density matrices of subsystems in 1D \cite{Eisler2006}. Recently, it has been shown that a large finite effective temperature can be obtained by considering lower-dimensional subsystems of a $D$-dimensional gapped Dirac fermion vacuum \cite{Moghaddam2022}}. This can be understood as a consequence of the gradient of the Fermi velocity vanishing, reducing (gravitational) acceleration to zero. However, we can also make it finite in the thermodynamic limit by choosing the scaling as $\lambda = N/2 -1 $. The temperature then becomes fixed and is constant for all system sizes. This gives a clear difference between the thermal spectra in the lattice and spacetime models.

Notice that the derivation of the thermal spectrum in the lattice model made no use of the part of the system to the left of the horizon. Since the hopping strictly goes to zero at $x=0$, the lattice separates into independent subsystems that do not influence one another after the quench. This same quality can be found in general relativity, where the temperature of Hawking radiation depends only on the position of the horizon, which is given by the mass of the black hole \footnote{As long as the black hole does not rotate and has no charge} and not on any details of the interior. Just as in general relativity, however, the part of the system behind the horizon must exist for a thermal spectrum to emerge. Whereas in gravitational theories the thermal spectrum may arise from a highly entangled thermofield double state, in the lattice theory we found that the quenched Hamiltonian needs to equal the entanglement Hamiltonian for the right half of the constant hopping model in order for the system to thermalize. This condition implies that the presence of a generic horizon in the quenched system does {\it not} suffice to ensure a purely thermal spectrum to emerge: if the quenched Hamiltonian contains a horizon but is different from the entanglement Hamiltonian, the resulting overlap matrix will be diagonal, but not proportional to the Fermi-Dirac distribution.
Conversely, however, starting from any initial Hamiltonian, and quenching to a Hamiltonian that is proportional to the entanglement Hamiltonian defined on one side of a horizon, causes the excitation spectrum to be perfectly thermal. Translating this to a gravitational system, the significance of the linear variation of the hopping indicates the importance of having a (local) Rindler metric. Only when the metric is locally Rindler (as opposed to proper coordinates of general non-constant acceleration) will we see purely thermal radiation.  

The simplified theoretical model introduced here, employing a quench between two static Hamiltonians, is a minimal model in which the emergence of temperature can be studied without being influenced by the dynamics of the horizon formation. It directly mimics the original Unruh and Hawking effects which follow from an instantaneous coordinate transformation.

Finally, the thermalization by synthetic horizons may be realized in a variety of existing experimental setups including controllable electronic systems, fermionized 1D spin chains, ultracold atoms or optical experiments. 
In all these setups, the only two crucial criteria to observe thermalization are the system being effectively described by free lattice fermions and the tunability of hopping (local coupling) parameters to engineer position-dependence and thus the horizon.
This can open a venue for exploring fundamental quantum-mechanical aspects alongside gravity and curved spacetimes in various condensed matter settings.

\section*{Acknowledgments} 
We thank Cosma Fulga, Flavio Nogueira and Viktor K\"onye for stimulating discussions and we thank Ulrike Nitzsche for technical assistance. We acknowledge financial support from the Prins Bernhard Cultuurfonds and the Deutsche Forschungsgemeinschaft (DFG, German Research Foundation), through SFB 1143 project A5 and the W\"urzburg-Dresden Cluster of Excellence on Complexity and Topology in Quantum
Matter- ct.qmat (EXC 2147, Project Id No. 390858490). A.G.M. acknowledges partial financial support from Iran
Science Elites Federation under Grant No. 11/66332 and from Academy of Finland under the Project No. 331094.

\renewcommand{\vec}{\mathbf}

\clearpage

\onecolumngrid

\appendix
\counterwithin{figure}{section}

\section{General relativity}
\label{sec:grav}
This section reproduces some well-known results in general relativity to allow for easy comparison to the condensed matter analogues in the main text. We construct the set of quantized field operators on top of both a flat and a particular curved spacetime. In addition, we find a thermal spectrum of excitations related to thermal radiation around horizons. 

\subsection{Flat spacetime}
\label{sec:FlatSpacetime}
Starting with flat, 1+1D Minkowski space, with metric 
\begin{align}
    ds^2 = dT^2 - dX^2,
    \label{eq:metricflat}
\end{align}
the geodesics of light are given by $X(T) = T + X_0$ for $c=1$. Treating the metric as a static background, we can define a spinor field by solving the Dirac equation
\begin{align*}
    (i \gamma^\mu \partial_\mu - m)\psi(X,T) = 0 
\end{align*}
where $\gamma^\mu$ are the gamma matrices and satisfy the algebra
$\{ \gamma^\mu, \gamma^\nu \} = 2\eta^{\mu\nu} I$. We choose the chiral (Weyl) representation of the Dirac $\gamma$ matrices, and in particular $\gamma_0 = i\sigma_x$ and $\gamma_1 = \sigma_y$. This Dirac equation can be used to define a Hamiltonian as there is a time-like killing vector \cite{Rodriguez2017}. With $\hat{p} = -i \partial_x$ and $i\partial_t \psi(X,T) = H \psi(X,T)$ we obtain $\hat{H} = \sigma_z \hat{p}.$

For massless fields, $m=0$, we have separate equations for $
\Psi^{\text{M}}_{E_\text{M}}(X,T) = e^{-i E_{\text{M}} T}[\psi_{E_\text{M}}^+(X), \psi_{E_\text{M}}^-(X)]$ that become 
\begin{align*}
    \pm iE_\text{M} \psi^\pm_{E_\text{M}}(X) & = \partial_X \psi^\pm_{E_\text{M}}(X)
\end{align*}
and the eigenfunctions are the left and right moving plane waves with energy ${E_\text{M}}$: 
\begin{align}
    \psi_{E_\text{M}}(X) = \frac{1}{\sqrt{2\pi}}[e^{i\epsilon_{\text{M}} X}, e^{-i\epsilon_{\text{M}} X}]. 
\end{align}
The factor $\frac{1}{\sqrt{2\pi}}$ comes from the normalization condition for a Dirac field~\cite{collas_klein_2019, parker_toms_2009},  
\begin{align*}
    (\psi_\epsilon, \psi_{\epsilon'}) = \int dV_X \psi_\epsilon^\dagger(X,T) \psi_{\epsilon'}(X,T) = \delta(\epsilon-\epsilon'). 
\end{align*} 
The field can be quantized by writing it as~\cite{SC}
\begin{align*}
    \hat{\Psi}^M(T, X) = \int_{-\infty}^{\infty} \frac{dk}{\sqrt{2 \pi }}\left(e^{-i|k|T+ikX}\hat{c}_k + e^{i|k|T-ikX}\hat{c}_k^\dagger \right)
\end{align*}
where $k$ ranges from $-\infty$ to $\infty$ and $E_{\text{M}} = |k|$. Notice that we defined positive and negative energy modes here with respect to the proper time of a static observer in the flat spacetime. This ensures that the particles $\hat{c}_k$ are the ones detected by a particle detector at a fixed position $X$ in flat spacetime.

\subsection{Curved spacetime}
\label{sec:CurvedSpacetime}
We need a spacetime with at least a horizon present to describe thermal radiation. A well-known metric that has this feature is 
\begin{align}
    ds^2 = \Gamma^2 x^2 dt^2 - dx^2.
    \label{eq:metric2}
\end{align}
for $x \geq 0$. We will leave the metric for $x<0$ undefined. The geodesics of light on the curved background are 
\begin{align}
    x(t) = x_0 e^{ \pm t \Gamma} 
    \label{eq:GeodesicsRind}
\end{align}
The light hits $x=0$ only at $t=\pm \infty$: a static observer cannot send or receive signals beyond $x=0$, and this point thus constitutes a horizon.

As before, a spinor field can be considered on the background described by this metric by solving the Dirac equation. We will do this for the more general metric $ds^2 = w(x) ( v(x)^2 dt^2 - dx^2) $. First, to make the Dirac equation covariant we need to replace the partial derivative with a covariant derivative and also make the gamma matrices position dependent so that they satisfy 
\begin{align*}
      \{ \Bar{\gamma}^\mu , \Bar{\gamma}^\nu \} = 2 g^{\mu\nu} I
\end{align*}
where the bar implicates a spatial dependence \cite{collas_klein_2019}. This gives the generally covariant Dirac equation 
\begin{align*}
    (i \Bar{\gamma}^\mu \nabla_\mu - m I) \Phi^\text{R}(\vec{x}, t) = 0.
\end{align*}
In 1+1D the Dirac equation simplifies to
\begin{align*}
    \Big[i\gamma^A e_A^\mu \partial_\mu + \frac{i}{2}\gamma^A \frac{1}{\sqrt{-g}}\partial_\mu(\sqrt{-g}e_A^\mu)-mI_2\Big]\Phi^\text{R}(x,t)=0
\end{align*} 
where $e$ is the zweibein. The zweibein has a vector label A running over $0,1$ and is defined by the relations 
\begin{align*}
    g_{\alpha \beta} = e_\alpha^A e_\beta^B \eta_{AB} \\
    g^{\alpha \beta} = e^\alpha_A e^\beta_B \eta^{AB}.
\end{align*}
As before, we choose the chiral (Weyl) representation of the Dirac $\gamma$ matrices, in particular we choose $\gamma_0 = -i \sigma_x$ and $\gamma_1 = \sigma_y$. 
The inverse of the metric is 
\begin{align*}
    & g^{00} = \frac{1}{v(x)^2 w(x)}\\
    & g^{11} = \frac{1}{w(x)}
\end{align*}
such that the zweibein becomes
\begin{align*}
    & e_0 =  \frac{1}{\sqrt{w(x)}|v(x)|}\partial_t \\
    & e_1 = \frac{1}{\sqrt{w(x)}} \partial_x.
\end{align*}
and the determinant $\sqrt{-g} = \sqrt{w(x)^2 v(x)^2}$.
With all of these, the Dirac equation becomes 
\begin{align}
    0 = & \left(  \frac{\sigma_x}{\sqrt{w(x)}|v(x)|}\partial_t + \frac{1}{\sqrt{w(x)}}i \sigma_y \partial_x \right. \nonumber \\ 
     + & \left.\frac{i\sigma_y}{2|w(x)v(x)|}\partial_x\left(\frac{|w(x)v(x)|}{\sqrt{w(x)}}\right) + m I \right) \Psi(x,t) 
    \label{eq:diracW}
\end{align}
Again we can use this to define a Hamiltonian for the massless fermions, yielding $\hat{H} =\sigma_z \Gamma ( \hat{x} \hat{p} +\hat{p}\hat{x})/2$.

For massless fields we have again separate equations for the components of the spinor $\Phi^{\text{R}}_{E_\text{R}}(x,t) = e^{-iE_\text{R} t}[\phi_{E_\text{R}}^+(x), \phi_{E_\text{R}}^-(x)]$ that become 
\begin{align*}
    \pm iE_\text{R} \phi^{\pm}_{E_\text{R}}(x)  = \sigma_z & \Big( |v(x)| \partial_x \\ 
    & +\frac{|v'(x)|}{2} + \frac{w'(x)v(x)}{4w(x)}\Big) \phi^{\pm}_{E_\text{R}}(x) 
\end{align*}
or 
\begin{align*}
    \phi^\pm_{E_\text{R}}(x) & = \phi_0 
    \exp\big({\int \frac{ \pm iE_\text{R}}{|v(x)|} -\frac{|v'(x)|}{2|v(x)|} -\frac{w'(x)}{4w(x)} dx}\big).
\end{align*}

Specialising to the case $|v(x)| = \Gamma |x|$, and considering fields that are defined for $x>0$ only, we find
\begin{align}
        \phi^\pm_{E_\text{R}}(x) & = \phi_0|x|^{\pm i E_\text{R}/\Gamma-1/2} \label{eq:psigravW}
\end{align}
Normalising with use of the relation $\int_0^\infty dx \, x^{i{E_{\text{R}}} -1} = 2\pi \delta({E_{\text{R}}})$ gives $\phi^\pm_0 = \sqrt{{1}/(4\pi)}$.
To quantise the particles we should define positive and negative frequencies with respect to a future directed timelike Killing vector \cite{carroll_2020}. On opposite sides of the horizon at $x=0$ this has a different sign (infalling waves move into opposite directions). We therefore have to define support of the wavefunctions differently on both sides. Since we did not define the metric for $x<0$ we will also leave the wave functions unspecified in that regime, and denote them by a general function $h_q(x,t)$. The quantum field then becomes 
\begin{align*}
    \hat{\Psi}^{\text{R}}(t,x) & = \int_{-\infty}^\infty \frac{dq}{\sqrt{4\pi}} \nonumber \\ 
    & \left(e^{-i|q|t} |x|^{iq/\Gamma -1/2} \hat{b}_q^{(1)} 
    + e^{i|q|t} |x|^{-iq/\Gamma-1/2} \hat{b}^{(1)\dagger}_q  \right. \\
    & \left. + h_q(x,t)\hat{b}_q^{(2)} 
    + h^*_q(x,t) \hat{b}^{(2)\dagger}_q
    \right).
\end{align*}
Here, the dispersion relation is defined as $\epsilon_{\text{R}} = |q|$. The operators $\hat{b}_k^{(1)}$ are defined such that they have support only on the region of space where $x>0$ and likewise $\hat{b}_k^{(2)}$ has support only on $x<0$. This implies that a particle detector at a fixed position $x>0$ in this curved spacetime will detect particles $\hat{b}_q^{(1)}$. 

\subsection{Thermal spectrum}
\label{ThermalSpectrum}
Returning to flat spacetime, consider the vacuum $\ket{0_{\text{M}}}$, defined as the state with all negative energy modes filled and positive energy modes empty.  We then have $\bra{0_{\text{M}}}c_k^\dagger c_{k'}^{\phantom\dagger}\ket{0_{\text{M}}} = \delta(k-k')\Theta(-k) $, with $\Theta(-k)$ the Heaviside step function. The $\hat{c}_k^\dagger$ operators are the creation operators for particles registered by a static detector in flat spacetime. 

Instantaneously quenching to the metric of equation~\eqref{eq:metric2}, a detector that is static with respect to the post-quench metric will register particles created by $\hat{b}_l^{(1)\dagger}$. Since an instantaneous quench should not affect the state of a system, the detector will register expectation values $\bra{0_{\text{M}}}\hat{b}_l^{(1)\dagger} \hat{b}^{(1)}_m\ket{0_{\text{M}}} $. We can calculate these using the (local) transformation
\begin{align*}
    \hat{b}^{(1)}_q = \int dk  \Bar{C}_{k, q} \hat{c}_k + C_{k,q} \hat{c}_k^\dagger.
\end{align*}
The vacuum expectation value is then written as
\begin{align*}
    \bra{0_{\text{M}}}\hat{b}^{(1)\dagger}_l \hat{b}^{(1)}_{m}\ket{0_{\text{M}}} & = \bra{0_{\text{M}}}  \left( \int dk C_{k,l}^* \hat{c}_k + \Bar{C}_{k, l}^* \hat{c}_k^\dagger\right) \\
    \times & \left(\int dk' C_{k',m} \hat{c}_{k'}^\dagger + \Bar{C}_{k', m} \hat{c}_{k'}\right)\ket{0_{\text{M}}}.
\end{align*}
Using the definition of the vacuum and the anticommutation relations $\{ \hat{c}_k^\dagger, \hat{c}_{k'}^{\phantom\dagger} \} = \delta_{k, k'} $ we are left with 
\begin{align*}
    \bra{0_{\text{M}}}\hat{b}^{(1)\dagger}_l \hat{b}^{(1)}_m\ket{0_{\text{M}}} = \int_{0}^\infty dk \Big( C_{k,l}^* C_{k, m} + \Bar{C}_{-k, l}^* \Bar{C}_{-k,m}\Big).
\end{align*}

The function $C$ can be calculated from the overlap of the wave functions defined on both metrics, using the orthonormality conditions. If we write the quantum fields as 
\begin{align*}
    \hat{\Psi}^{\text{M}}(t, x) & = \int_{-\infty}^{\infty} dk f_k \hat{c}_k + f_k^*\hat{c}_k^\dagger \\
    \hat{\Psi}^{\text{R}}(t,x) & = \int_{-\infty}^{\infty} dk g_k \hat{b}_k + g_k^*\hat{b}_k^\dagger
\end{align*}
then $C$ can be expressed as \cite{carroll_2020}
\begin{align*}
    C_{k, q} & = -(g_q, f_k^*) \notag \\&
    =  -\int_0^\infty \frac{dx}{2\sqrt{2}\pi} e^{i|q|t}x^{-iq/\Gamma-1/2} e^{i|k|T-ikX} \\
    \Bar{C}_{k, q} &= (g_q, f_k) = \int_0^\infty \frac{dx}{2\sqrt{2}\pi} e^{i|q|t}x^{-iq/\Gamma-1/2} e^{-i|k|T+ikX}.
\end{align*}
These relations constitute a Bogoliubov transformation relating the modes in different spacetimes. To evaluate these expressions further, one needs to relate the coordinates $T, X$ to $t, x$. When switching from an inertial (static in flat spacetime) to an accelerated observer (static in curved spacetime), we have the relations 
\begin{align*}
    T &= x \sinh(\Gamma t) \\
    X &= x \cosh(\Gamma t).
\end{align*}
However, as the quench happens instantaneously at $t=0$, {we insist that the coordinates agree throughout the time slice $t=T=0$}. In contrast to previous research on synthetic black holes, this means the quench we consider here is not a switch between two distinct observers. Rather, we consider a single observer that instantaneously starts accelerating with acceleration $\alpha = \frac{1}{\Gamma x_0}$. Or, by use of the equivalence principle: the instantaneous creation of a black hole with a horizon at $x=0$. 

The overlap function $C$ can now be seen as the Fourier components of the curved spacetime wave function $\psi_{\text{R}}(x)$, given by  
\begin{align}
     C_{k, q} = \frac{1}{2\sqrt{2}\pi} & e^{-\pi/2 \text{sgn}(k)(q/\Gamma+i/2) \nonumber} \\ 
     \times & |k|^{(iq/\Gamma-1/2)} \Gamma\left(-iq/\Gamma+1/2\right)
     \label{eq:C0}.
\end{align}
Using the transformation $u=ikx$ and the definition $\Gamma(z) = \int_0^{\pm i \infty} du \, u^{z-1}e^{-u}$, this becomes
\begin{align*}
    C_{k, q} 
    & = \frac{1}{2\sqrt{2}\pi} e^{i(|q|+|k|)t} \int_0^{i\infty} \frac{du \, u^{(-iq/\Gamma+1/2)-1}}{(ik)(ik)^{-iq/\Gamma-1/2}}  e^{-u} \\
    & = \frac{1}{2\pi} e^{i(|q|+|k|)t} (ik)^{iq/\Gamma-1/2} \Gamma\left(-iq/\Gamma +1/2)\right). 
\end{align*}
Next we can choose to write $\pm i = e^{\pm i \pi/2}$ such that we end up with 
 \begin{align}
     C_{k, q}=\frac{1}{2\sqrt{2}\pi} & e^{i(|q|+|k|)t} e^{-\pi/2 \text{sgn}(k)(q/\Gamma+i/2) \nonumber} \\ 
     \times & |k|^{(iq/\Gamma-1/2)} \Gamma\left(-iq/\Gamma+1/2\right).
 \end{align} Similarly we get 
 \begin{align*}
     \Bar{C}_{k, q} = -\frac{1}{2\sqrt{2}\pi} & e^{i(|q|-|k|)t} e^{\pi/2 \text{sgn}(k)(q/\Gamma+i/2)\nonumber} \\ 
     \times & |k|^{(iq/\Gamma-1/2)} \Gamma\left(-iq/\Gamma+1/2\right).
 \end{align*}
Integrating $C$ to finally find the expectation value, gives 
 \begin{align*}
     \int_{0}^{\infty} C_{k, q}^* C_{k, q'} dk  & = \frac{1}{8\pi^2}e^{i(|q|-|q'|)t} e^{-(q+q')\pi/(2\Gamma)} \Gamma\left(\frac{1}{2} -\frac{iq}{\Gamma}\right) \\ 
     & \times \Gamma\left(\frac{1}{2} + \frac{iq'}{\Gamma}\right) \int_0^{\infty} |k|^{i/\Gamma(q-q')-1} dk \\
     & = \frac{1}{4 \pi} e^{-q\pi/\Gamma} |\Gamma(-iq/\Gamma+1/2)|^2 \delta(q-q') \\
     & = \frac{1}{2} \frac{1}{1+ e^{2 \pi q /\Gamma}} \delta(q-q').
 \end{align*}
Here, we used $\int_0^\infty dx \, x^{iw-1} = 2\pi \delta(w)$ and $|\Gamma(1/2+ix)|^2 = \pi \sech(\pi x)$. For $\Bar{C}$ we likewise find
\begin{align*}
    \int_{0}^{\infty} \Bar{C}_{-k, q}^* \Bar{C}_{-k, q'} dk & = \int_{0}^\infty C_{k, q}^* C_{k, q'} dk = \\
     & = \frac{1}{2} \frac{1}{1+ e^{2 \pi q /\Gamma}} \delta(q-q').
\end{align*}
With these, the vacuum expectation value becomes 
\begin{align}
        \bra{0_{\text{M}}}\hat{b}^{(1)\dagger}_q \hat{b}^{(1)}_{q'}\ket{0_{\text{M}}}  =  \frac{1}{1+ e^{2 \pi q /\Gamma}} \delta(q-q'). \label{eq:FDGR}
\end{align}
This is the Fermi-Dirac distribution for particles in thermal equilibrium with temperature $T = {\Gamma}/(2\pi)$.

\section{Comparison lattice and spacetime}
\label{sec:AppComp}
In this section we give a more detailed comparing of the lattice eigenstates and the modes of their analogue gravitational models.

\subsection{Flat spacetime and its analogue}
First of all we need to consider the condensed matter analogue for flat spacetime, a tight binding lattice model with constant hopping $t_j=-\lambda$. 

To obtain the exact eigenfunctions of the Hamiltonian with finite $N$, we need to impose the constraints
\begin{align*}
    & \psi_{E_\text{M}}(N+1) = \psi_{E_\text{M}}(0) =  0 \\
    & \psi_{E_\text{M}}(j+1)+ \psi_{E_\text{M}} (j-1) = \frac{E_\text{M}}{\lambda} \psi_{E_\text{M}}(j)
\end{align*}
where the eigenfunctions $\psi_{E_\text{M}}(j)$ of the Hamiltonian are defined as $\psi_E = \sum_{j=1}^{N} \psi_E(j) c_j^\dagger$, expressed in the position basis. Making the assumption $\psi_{E_\text{M}}(j) \propto \sin(\theta j)$ all constraints can be satisfied by choosing $\theta = \pi l /(N+1) + m \pi $, for all $l,m \in \mathbb{Z}$. The energy spectrum then has the $N$ distinct values $E_\text{M}(l) = 2\lambda \cos(\pi l/(N+1) + m\pi)$ with $l \in {1,2, ...N}$ and $m \in \mathbb{Z}$. Defining $l=1$ to be the lowest (most negative) energy, we need $m$ to be odd. Choosing $m=-1$ then gives the eigenfunctions
\begin{align}
    \psi_l(j) = \sin(\frac{\pi l j}{N+1} - j \pi ) = (-1)^{j} \sin(\frac{\pi l j}{N+1} )
    \label{eq:MWFexact}
\end{align}
with energies $E_\text{M}(l) = -2\lambda \cos(\pi l/(N+1))$.

We can determine the semi-classical group velocity for a wave packet in the low energy limit. Considering a half-filled band with Fermi momentum $k_{\text{F}} = \frac{\pi l}{N+1}  = \pm \pi/2$, the dispersion is approximately linear around the Fermi energy: $E(k_{\text{F}} + k' ) \approx \lambda k' $. Here $k'$ is a small perturbation away from the Fermi momentum. Using the semi-classical continuum definition, this gives the trajectories for wave packets of $\langle x(t) \rangle = x_0 \pm \lambda t$ and a group velocity (and phase velocity) equal to $\lambda$ \cite{ali}. We recognize the linear behavior typical for trajectories of light with velocity $\lambda= c$ in flat spacetime. 

There are a few important differences between the gravitational model and its lattice analogue. First is range of the energies in the spectrum. In the lattice model, energies ranges from the finite values of $-2\lambda$ to $2\lambda$, due to the periodicity in $k$ which in turn is generated by the discreteness of the lattice. The gravitational model, on the other hand, is continuous in space and therefore has a spectrum extending from $-\infty$ to $\infty$. The analogy will therefore necessarily work only in the low energy limit. 

\begin{figure}[tb]
    \centering
    \includegraphics[width=\linewidth]{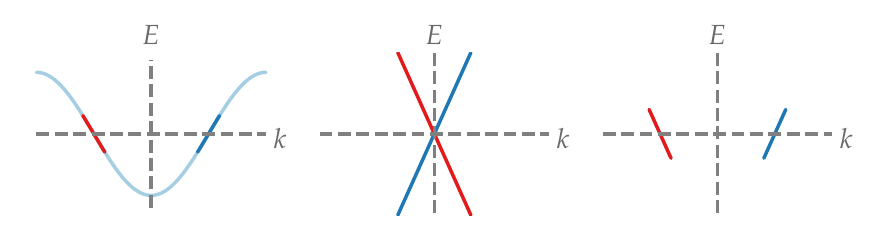}
    \caption{Dispersion relations for the lattice model (left), the gravitational model (middle) and the gauge shifted gravitational model (right)}
    \label{fig:Dis}
\end{figure}

Secondly, as illustrated in Fig. \ref{fig:Dis}, the low energy modes in the two models are situated at different values of $k$. 
The left frame shows the dispersion relation of the constant hopping lattice model with periodic boundary conditions, with low energy modes (i.e. energies close to $E_{\text{F}}$) situated around $k = \pm \pi/2$. The middle frame shows the dispersion relation of modes in the gravitational model of flat spacetime, with low energy modes situated around $k=0$. To have the low energy modes at comparable $k$-values we can implement a gauge transformation on the Dirac equation in the gravitational model, by considering a metric conformal to flat spacetime
\begin{align}
    ds^2 = e^{2 \pi i X}(dT^2 - dX^2). \label{eq:metricgauge}
\end{align}
For all integer values $X \in \mathcal{Z}$ the metric is unaffected. In between positions corresponding to lattice points, however, it is multiplied by a phase. This complex conformal factor does not alter the geodesics of light, but does change the Dirac equation to
\begin{align*}
    (i\gamma^u \partial_u - \gamma^0 \pi/2)\Psi^\text{M}(X,T) = 0.
\end{align*}
The wave functions then become 
\begin{align}
    \Psi^{\text{M}}_{E_\text{M}}(X,T) =& \frac{1}{\sqrt{2\pi}} e^{i|E_{\text{M}}| T} \nonumber \\ 
    \times & [e^{i(E_{\text{M}}+\pi/2)X}, e^{i(-E_{\text{M}}+\pi/2)X} ],
    \label{eq:GaugeWM}
\end{align}
as explained in detail in section \ref{sec:grav}. The dispersion is shifted to $E_{\text{M}} = k \mp \pi/2$.

\begin{figure}[tb]
    \centering
    \includegraphics[width=\linewidth]{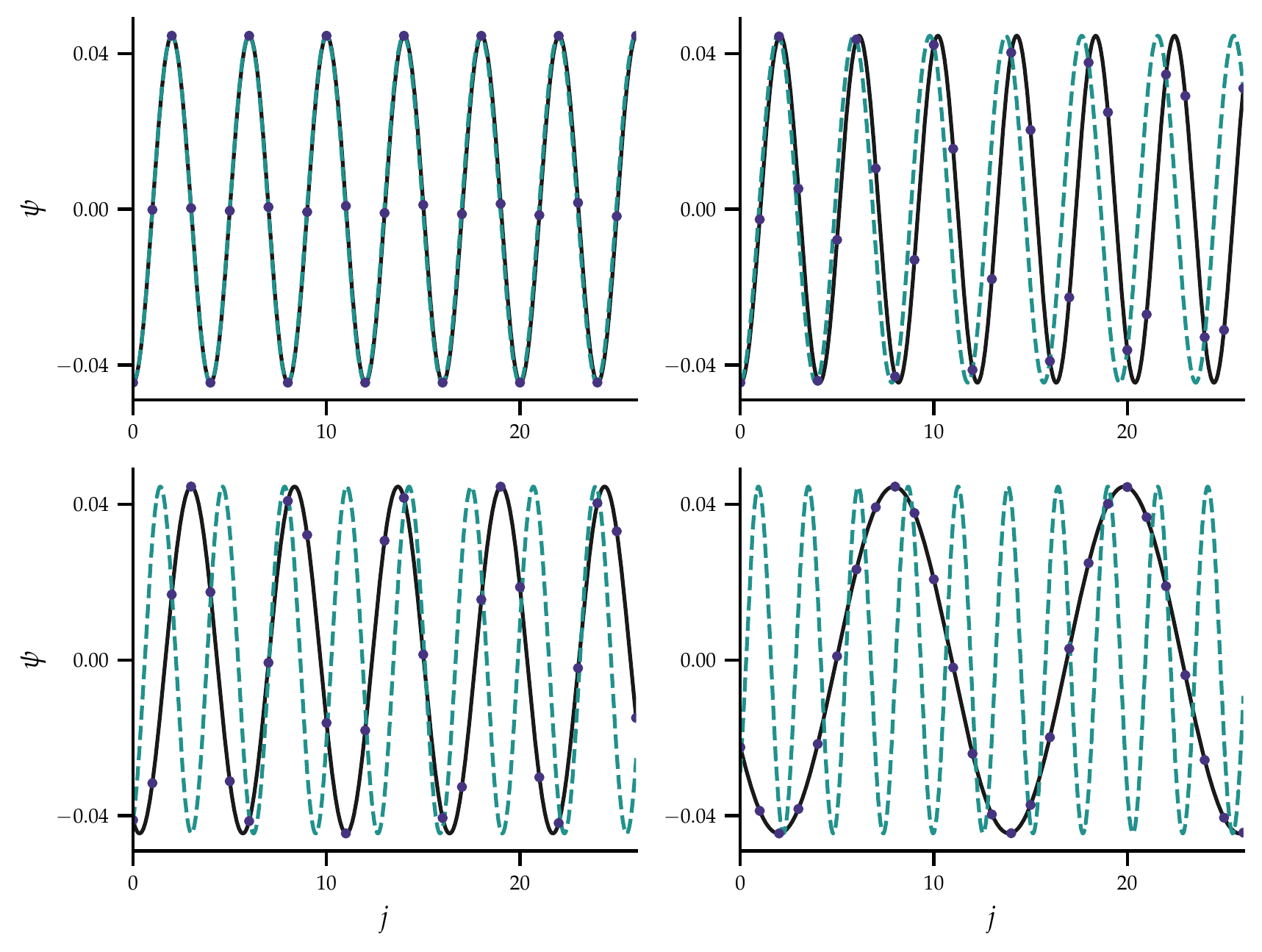}
    \caption{For a lattice of size $N=1006$ the first, 10th, 128th and 335th eigenfunctions above the zero energy level are plotted. The blue dots and black line are the exact solutions of equation \ref{eq:MWFexact}. The dashed blue line is the solution of equation \ref{eq:GaugeWM} normalised in the domain $X=0$ to $N$. }
    \label{fig:MWF4}
\end{figure}

Next, we can compare the wave functions of equation \eqref{eq:GaugeWM} with the lattice wave functions of equation~\eqref{eq:MWFexact}, as plotted in Fig.~\ref{fig:MWF4}. 
The blue dots represent the the wave functions of equation \eqref{eq:MWFexact}.
The real part of the gravitational wave functions of equation \eqref{eq:GaugeWM} are given by the green dashed line. The black line is the continuous version of the lattice wave functions by defining $X = j - N/2$ such that the middle of the lattice is at $X=0$, and by extending the solutions to all $X$ between $X=-N/2 +1 $ and $X=N/2 -1$. This continuous extension is drawn to better compare the gravitational and lattice solutions. The upper left panel shows the first energy level above zero ($E_{\text{F}}$), where all three wave functions overlap The upper right panel displays the 10th energy level above zero and the gravitational wave function can be seen to start diverging from the lattice wave functions. For even higher energies, as depicted in the lower two panels, the gravitational wave functions become even more different from their lattice analogues. This is expected as only close to zero energy the lattice model has approximately the same dispersion relation as the gravitational model.

Finally, notice that the gravitational wave functions are complex. Due to the open boundary conditions on the lattice, the wave functions for the lattice model instead are strictly real, thus diverging from the gravitational solution. 

\subsection{Curved spacetime and its analogue}
For curved spacetime the condensed matter analogue is the linear hopping tight binding model, the trajectories of a wave packet on this model match those on the Rindler metric. For the wavefunctions the same comparison between the lattice and gravitational models can be made. As before, a shift $k \to k \pm \pi/2$ is necessary for the lattice eigenfunction at $E_{\text{F}}$ to match the zero energy mode in the gravitational model. A factor $e^{-2\pi i x}$ in front of the metric again gives this shift and introduces a factor $e^{i \pi/2 x} $ in front of the wave functions such that they reduce to 
\begin{align}
    \frac{1}{\sqrt{4\pi}} e^{i \pi/2 x} |x|^{-1/2+i \epsilon/\Gamma}.
\end{align}
These agree with the low-energy approximate eigenfunctions 
\begin{align}
    \phi_n = \frac{1}{\sqrt{2\pi}} i^n n^{-1/2 + iE_R/(2\lambda)(N/2 -1)}
    \label{eq:approxlatticemodel}
\end{align}
for the linear hopping model for $\Gamma = 2\lambda/(N/2-1)$. 

\begin{figure}[tb]
    \centering
    \includegraphics[width=\linewidth]{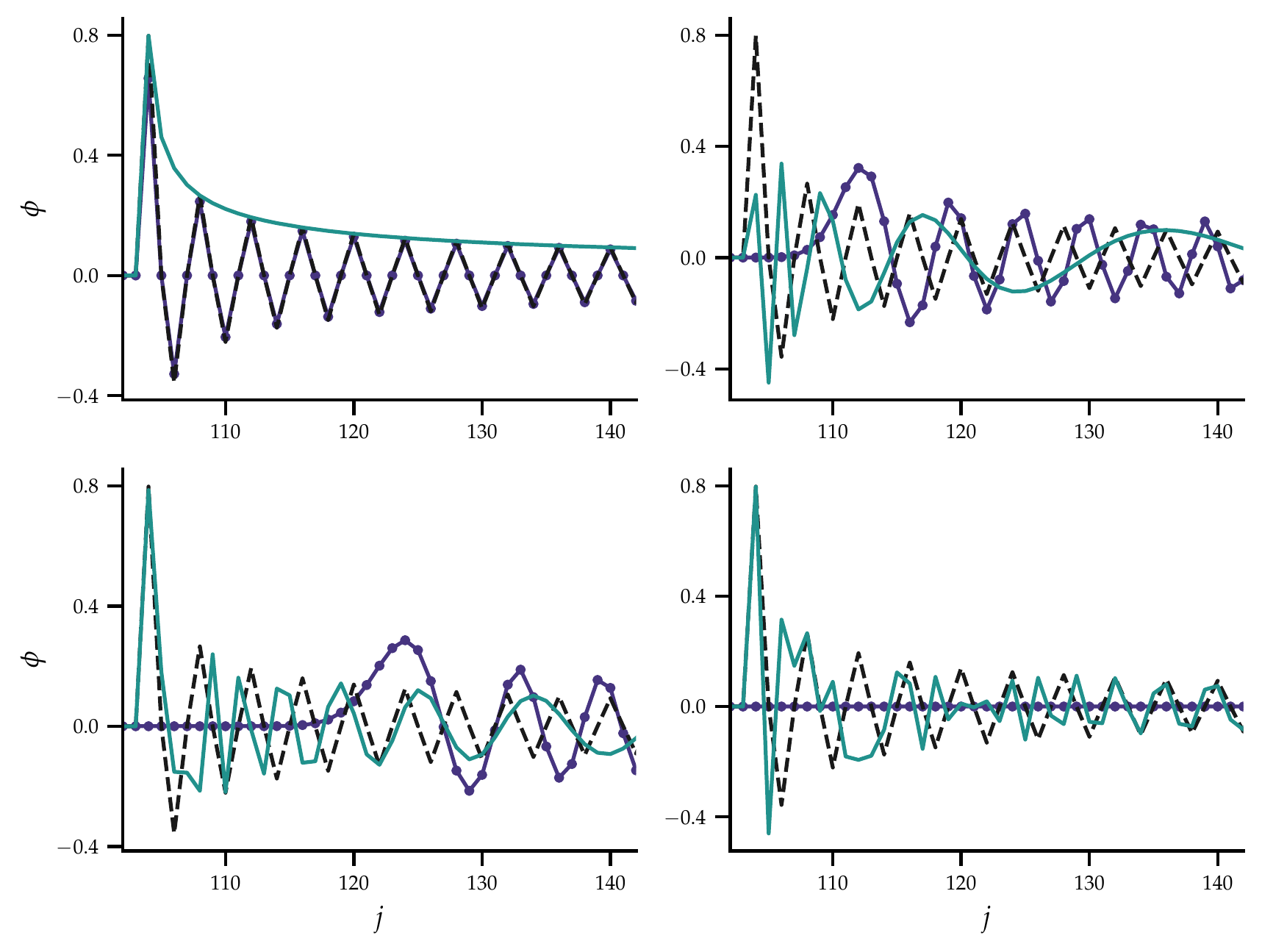}
    \caption{For a lattice of size N=1006, the first, 10th, 128th and 335th eigenfunctions are plotted. The blue dots and blue line are the exact eigenfunctions of the lattice Rindler Hamiltonian with $\gamma=1$. The green line is the gravitational solution with $\Gamma=2\lambda/(N/2-1)$ and the black dashed line is the continuous solution of equation \ref{eq:approxlatticemodel}. }
    \label{fig:WFR4}
\end{figure}

The lattice Hamiltonian can also be diagonalized numerically, as shown in Fig. \ref{fig:WFR4}. The black line indicates the continuous approximate solution of equation  \eqref{eq:approxlatticemodel}, while the blue dots and line show the numerical eigenfunctions of the lattice Hamiltonian. The green line is the exact solution to the gravitational model with $\Gamma=2\lambda/(N/2-1)$.
The gravitational modes of equation are an envelope to the exact solution to the lattice model at zero energy. Likewise, at low energies the low-energy approximation to the solutions of the lattice model can be seen to work. For non-zero energies the functions start to diverge and the exact numerical solution becomes zero in a region around $j=0$, while the continuous approximate form does not. 

\subsection{The thermal spectrum}
\begin{figure}[tb]
    \centering
    \includegraphics[width=\linewidth]{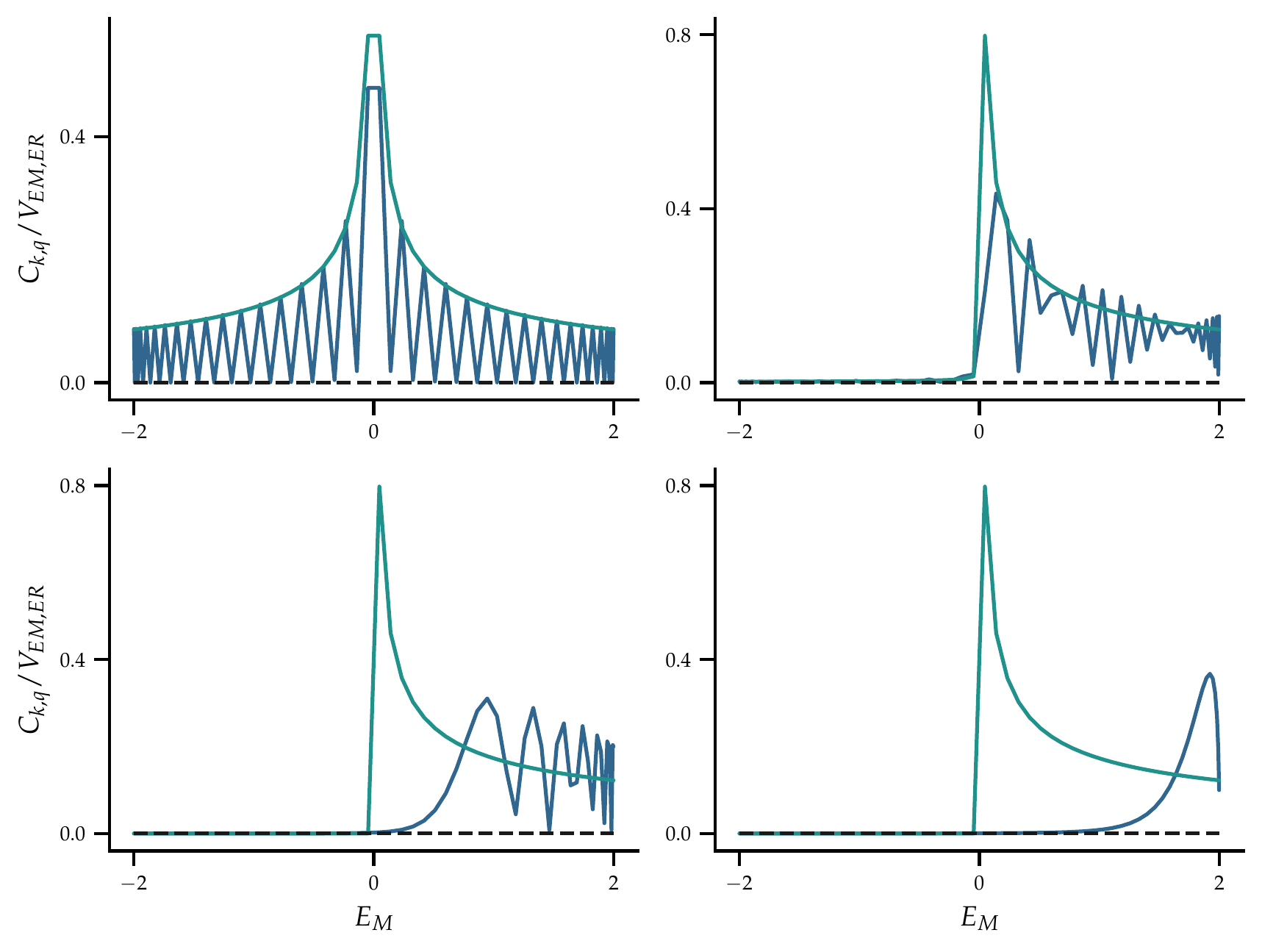}
    \caption{In blue, the absolute value of the numerical overlap function $V_{E_{\text{M}},E_{\text{R}}}$ defined on the lattice is plotted for different Rindler energies $E_{\text{R}},$. In green, the absolute value of the analytical expression for the corresponding gravitational overlap function $C_{k,q}$ is shown.}
    \label{fig:C2N}
\end{figure}

For both sets of gravitational modes (flat and curved spacetime, corresponding to before and after the quench) a conformal prefactor was required to ensure the correct correspondence to the eigenfunctions of the analogue lattice models. In the calculation of the overlap matrix, however, these two factors cancel and the result of the overlap matrix of general relativity in section \ref{sec:grav} can be compared directly to $V_{E_{\text{M}},E_{\text{R}}} = \bra{E_M}\ket{E_R}$, as shown in Fig. \ref{fig:C2N}.
The absolute value of the overlap function in the gravitational model ($|C|$, shown in green) for energy $E_{\text{R}}=0$ fits almost perfectly as an envelope to the corresponding overlap function of lattice model ($|V|$, indicated in blue). For nonzero energies $E_{\text{R}}$, the overlap functions reduce to zero on the same scale for negative energies $E_{\text{M}} < 0 $. They start to disagree, however, for $E_{\text{M}} > 0$, the maximum of $|C|$ always remaining at $E_{\text{M}} =0$ while the maximum of $V$ moves away towards the maximum value of $E_{\text{M}} = 2$. Similar behavior was found for the wave functions of Fig. \ref{fig:WFR4}, where for higher energies the weights of the wave functions in the lattice model vanish near the origin.

\section{The entanglement Hamiltonian}
\label{sec:AppEntan}
In this section, we reproduce the well-known result that the eigenstates of an entanglement Hamiltonian give rise to a thermal spectrum~\cite{Peschel2009}.

Consider the general case of a free fermionic system which can be divided into two parts, left (L) and right (R). If the full system is in thermal equilibrium, its density matrix in the canonical ensemble is given by $\rho = e^{-\beta H}$. The reduced density matrix describing the properties of subsystem R can be obtained by performing a trace over the left subsystem 
\begin{align*}
    \rho_{\text{R}} = \sum_{\{n_i\}=0,1}\Motimes_{i\in L } \bra{n_i}  \rho  \ket{n_i}
\end{align*}
or
\begin{align*}
    \rho_{\text{R}} = \sum_{\{n^{(L)}_{i}\}=0,1} \bra{n^{(L)}_{1},\cdots, n^{(L)}_{N'}}  \rho  \ket{n^{(L)}_{1},\cdots, n^{(L)}_{N'}}
\end{align*}
where $n_i$ is the occupation number for site $i$. 

The entanglement Hamiltonian $h_{\text{R}} = \log(\rho_{\text{R}})$ quantifies the entanglement between subsystems L and R. It is important to notice that despite its name, the entanglement Hamiltonian is thus not a Hamiltonian for the subsystem. Next, we use Wick's theorem, which states that correlation functions for a free-fermionic Hamiltonian can always be written as products of two-point correlators. As this holds for the whole system, it must also be true within the two subsystems L and R separately. We thus know that $\rho_{\text{R}}$ factorises and is defined by all its 2-point correlation functions. As $\rho_{\text{R}}$ is also uniquely defined, we can then use Wick's theorem in reverse and find that $h_{\text{R}}$ must in turn be a free fermionic system, as it can be defined entirely in terms of 2-point correlators~\cite{Peschel2009}. 

Writing the entanglement Hamiltonian in diagonal form thus gives 
\begin{align*}
    \rho_{\text{R}} = K e^{\left( -\sum_k \epsilon_k \hat{b}_k^\dagger \hat{b}_k^{\phantom\dagger}\right)}
\end{align*}
where the operators $\hat{b}_k$ are related to the operators $\hat{c}_j$ annihilating a particle at site $j$ as $\hat{c}_j = \sum_k \phi_{jk}\hat{b}_k$. These transformations define the real-space matrix elements of $h_{\text{R}}$ to be 
\begin{align}
\label{eq:Def_hR}
    h_{\text{R}}(i, j) = \sum_k \phi^*_{ki}\phi_{kj} \epsilon_k.
\end{align}

The correlation functions for the eigenstates of $h_{\text{R}}$ are given by
\begin{align*}
    \langle b_k^\dagger b_l^{\phantom\dagger} \rangle = \Tr \left( \rho_{\text{R}} b_k^\dagger b_l^{\phantom\dagger} \right) =  \Tr \left ( b_l^{\phantom\dagger} \rho_{\text{R}} b_k^\dagger  \right) ,
\end{align*}
where we used the cyclic property of the trace. If we pull $b_k^\dagger$ back through $\rho_{\text{R}}$ we get an extra factor from 
\begin{align}
    \rho_{\text{R}} b_k^\dagger = K e^{ - \sum_l \epsilon_l b_l^\dagger b_l^{\phantom\dagger} } b_k^\dagger =  b_k^\dagger \rho_{\text{R}} e^{-\epsilon_l}.
\end{align}
We thus find that
\begin{align*}
    \langle b_k^\dagger b_l^{\phantom\dagger} \rangle = \Tr(\rho_{\text{R}} b_k^\dagger b_l^{\phantom\dagger}) & = e^{-\epsilon_k} \Tr( b_l^{\phantom\dagger} b_k^\dagger \rho_{\text{R}}) \notag \\ 
    &= e^{-\epsilon_k} \Tr( (\delta_{l, k} - b_k^\dagger b_l^{\phantom\dagger}) \rho_{\text{R}} ) \\
    & = e^{-\epsilon_k} \left( \delta_{l,k} - \langle b_k^\dagger b_l^{\phantom\dagger}\rangle  \right) 
\end{align*}
In the third line we assumed normalization $\Tr(\rho_{\text{R}}) = 1$ and used the cyclic property of the trace again to arrive at the definition of the correlation function. Rearranging, we end up with the thermal distribution
\begin{align}
\label{eq:correlations_a}
    \langle b_k^\dagger b_l^{\phantom\dagger} \rangle & = \frac{\delta_{k, l}} {e^{\epsilon_k} +1 },
\end{align}
where $\epsilon_k$ are the eigenvalues of the entanglement Hamiltonian~\cite{Peschel2009}.

\end{document}